\documentclass[conference]{IEEEtran}
\IEEEoverridecommandlockouts
\usepackage{cite}
\usepackage{amsmath,amssymb,amsfonts}
\usepackage{algorithmic}
\usepackage{graphicx}
\usepackage{textcomp}
\usepackage{xcolor}
\usepackage[english]{babel}
\usepackage{siunitx}

\usepackage{cleveref}
\usepackage[sharp]{easylist}
\usepackage{enumitem}  
\usepackage{float}
\usepackage{minted}
\usepackage{sidecap}
\sidecaptionvpos{figure}{t}
\usepackage{subcaption}
\usepackage{todonotes}
\usepackage[normalem]{ulem} 
\usepackage{url}
\usepackage{xspace}
\usepackage{booktabs}

\usepackage{listings}
\lstdefinestyle{SQLwithhints}{
    language=SQL,
    morekeywords={with,hint,generated,simulators,ensemble,method,weights,real,register,simulator,executable,parameters,output_format,depends,geometry,references} 
}

\usepackage{color}
\definecolor{schalkeblau}{RGB}{0,0,204} 

\newcommand{\smartdb}{\textsc{G\MakeLowercase{en}IE}\xspace}

\def\BibTeX{{\rm B\kern-.05em{\sc i\kern-.025em b}\kern-.08em
    T\kern-.1667em\lower.7ex\hbox{E}\kern-.125emX}}
\begin{document}

\title{\smartdb: Simulator-Driven Iterative Data Exploration for Scientific Discovery
}

\author{
\IEEEauthorblockN{
Ashwin Gerard Colaco\IEEEauthorrefmark{1},
Martin Boissier\IEEEauthorrefmark{2},
Sriram Rao\IEEEauthorrefmark{1},
Shubharoop Ghosh\IEEEauthorrefmark{3},
Sharad Mehrotra\IEEEauthorrefmark{1},
Tilmann Rabl\IEEEauthorrefmark{2}
}
\IEEEauthorblockA{\IEEEauthorrefmark{1}University of California, Irvine, USA\\
Email: \{acolaco, sharad, srirar1\}@uci.edu}
\IEEEauthorblockA{\IEEEauthorrefmark{2}Hasso Plattner Institute, Germany\\
Email: \{martin.boissier, tilmann.rabl\}@hpi.de}
\IEEEauthorblockA{\IEEEauthorrefmark{3}ImageCat Inc, USA\\
Email: sg@imagecatinc.com}
}

\maketitle

\begin{abstract}
Physics-based simulators play a critical role in scientific discovery and risk assessment, enabling what-if analyses for events like wildfires and hurricanes. Today, databases treat these simulators as external pre-processing steps. Analysts must manually run a simulation, export the results, and load them into a database before analysis can begin. This linear workflow is inefficient, incurs high latency, and hinders interactive exploration, especially when the analysis itself dictates the need for new or refined simulation data.

We envision a new database paradigm, entitled \smartdb, that seamlessly integrates multiple simulators into databases to enable dynamic orchestration of simulation workflows. By making the database "simulation-aware," \smartdb can dynamically invoke simulators with appropriate parameters based on the user's query and analytical needs. This tight integration allows \smartdb to avoid generating data irrelevant to the analysis, reuse previously generated data, and support iterative, incremental analysis where results are progressively refined at interactive speeds.

We present our vision for \smartdb, designed as an extension to PostgreSQL, and demonstrate its potential benefits through comprehensive use cases: wildfire smoke dispersion analysis using WRF-SFIRE and HYSPLIT, and hurricane hazard assessment integrating wind, surge, and flood models. Our preliminary experiments show how \smartdb can transform these slow, static analyses into interactive explorations by intelligently managing the trade-off between simulation accuracy and runtime across multiple integrated simulators. We conclude by highlighting the challenges and opportunities ahead in realizing the full vision of \smartdb as a cornerstone for next-generation scientific data analysis.
\end{abstract}


\section{Introduction}

\begin{figure}[]   
    \begin{subfigure}[t]{\columnwidth}
        \begin{minipage}[c]{0.05\textwidth}
        \caption{}
        \end{minipage}
        \begin{minipage}[c]{0.95\textwidth}
        \centering
        \includegraphics[width=0.77\textwidth]
        {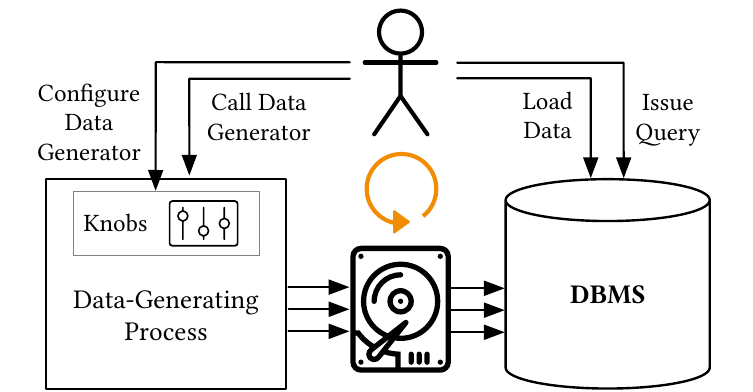}
        \end{minipage}\hfill
        \label{fig:intro_plot:before}
    \end{subfigure}
    \hfill
    \vspace{5mm}
    \begin{subfigure}[t]{\columnwidth}
        \begin{minipage}[c]{0.05\textwidth}
        \caption{}
        \end{minipage}
        \begin{minipage}[c]{0.94\textwidth}
        \includegraphics[width=\textwidth]
        {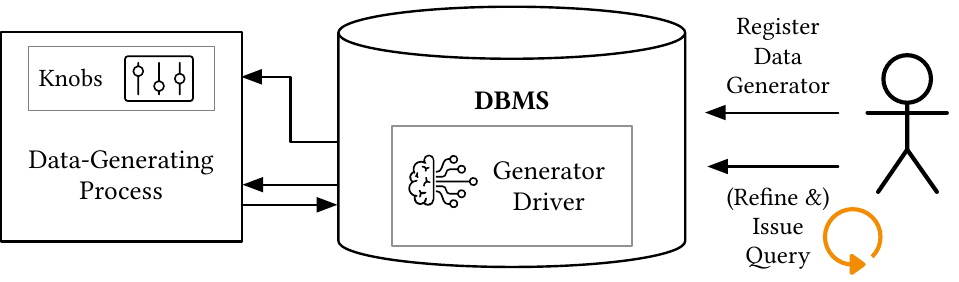}
        \label{fig:intro_plot:after}
        \end{minipage}\hfill
        
    \end{subfigure}
    
    \caption{Transformation from current static workflow to \smartdb's integrated approach. (a) Current workflow: Users manually configure simulators, wait for completion, export results, load into database, and analyze -- a repetitive cycle that must restart for any parameter change. (b) \smartdb workflow: Users register simulators once, then query their output directly via SQL; \smartdb automatically orchestrates simulator execution based on query needs.}
    \label{fig:intro_plot}
\end{figure}

Despite significant developments in database technologies, many scientific applications that rely on data from physics-based simulators remain bottlenecked by a linear, static workflow that severely limits what-if analysis capabilities. What-if analysis is commonly needed in exploratory data analysis tasks and planning scenarios, where analysts must evaluate multiple hypothetical situations, parameter variations, or alternative strategies. Today's approaches prevent interactive usage and make it nearly impossible to pose follow-up questions without restarting entire computational workflows~\cite{DBLP:journals/cacm/AdamSZTLSMCSKGM13}.

The problem becomes even more complex when multiple simulators must be integrated to answer real-world analytical questions. Consider simulation-driven consequence analysis, used to estimate the potential damage from natural disasters like wildfires or hurricanes~\cite{NOAAsHYSPLITAtmosphericTransportandDispersionModelingSystem, DescriptionandVerificationoftheNOAASmokeForecastingSystemThe2007FireSeason}. For wildfire analysis, forest managers need fire spread simulators like WRF-SFIRE to predict fire boundaries, atmospheric transport models like HYSPLIT to predict smoke dispersion, and health impact models to estimate population exposure. Similarly, hurricane hazard assessment requires integrating wind field models, storm surge simulators, and inland flood models. 

\Cref{fig:intro_plot}(a) illustrates the typical workflow for such analyses: a user manually configures each simulator with appropriate parameters and input files, waits for execution to complete (often hours or days), exports the output files in simulator-specific formats, loads the data into a database, and finally correlates results across different simulation outputs. When analytical needs change -- discovering that finer spatial resolution is needed in one area, or that an additional time period must be simulated -- the entire process must restart from the beginning. This static and repetitive process is inefficient and cumbersome, requiring analysts to predict their analytical needs upfront and preventing interactive exploration when findings from one simulation might dictate the parameters needed for another.

\paragraph*{The Static Workflow Problem}
The fundamental limitations of current approaches manifest in several critical ways. First, sequential execution bottlenecks force analysts to run simulators in a fixed order, where each step must complete entirely before the next can begin. For complex multi-simulator workflows, this leads to extremely long end-to-end execution times that can span days or weeks. Second, the inability to adapt parameters based on intermediate results leads to either over-generation of data (running simulations at maximum resolution for all possible areas, wasting computational resources) or under-generation (facing long delays when analysis requires more detail than initially anticipated). Third, redundant computation across similar scenarios wastes resources. When hurricane track predictions are similar across consecutive forecast updates, for instance, hazard calculations could be partially reused, but current systems cannot identify these opportunities. Most critically, when analytical needs change dynamically based on intermediate results, the static approach requires completely restarting workflows, making iterative scientific discovery impractical.

\paragraph*{The \smartdb Vision}
We envision a system, entitled \emph{\bf Gen}erator-Driven \emph{\bf I}terative Data \emph{\bf E}xploration (\smartdb), that overcomes these issues by seamlessly integrating multiple simulators into databases and enabling dynamic orchestration of their execution. \Cref{fig:intro_plot}(b) depicts the \smartdb approach: users register simulators once with the database system, defining their interfaces and parameter spaces, then issue SQL queries that reference simulation results directly. \smartdb automatically determines when to invoke which simulators, at what parameter settings, and for what spatial or temporal extent -- all driven by the analytical needs expressed in queries. This transforms physics-based simulators from isolated computational tools into first-class database objects that can be queried, composed, and optimized like traditional database operations.

The key insight behind \smartdb is to make databases "simulation-aware" by modeling simulators as configurable gray boxes. While most of their internal logic remains encapsulated, simulators expose APIs and parameters that control execution characteristics such as spatial resolution, temporal resolution, Monte Carlo iterations, convergence thresholds, and domain boundaries. These control parameters enable \smartdb to dynamically steer simulator execution to meet end-user analytical tasks at interactive speeds. Rather than requiring users to master the complexities of each simulator's configuration, \smartdb handles the translation from high-level analytical requirements (expressed in SQL) to low-level simulator parameters.

This paper makes the following contributions:
\begin{itemize}
    \item We introduce a data model for simulator-generated virtual attributes that extends relational databases with physics-based simulators as first-class computational components.
    \item We present the \smartdb architecture, describing how query-driven orchestration, state management, and simulator adapters enable interactive what-if analysis.
    \item We demonstrate \smartdb's applicability through wildfire smoke dispersion and hurricane hazard assessment, showing how multiple simulators can be composed.
    \item We provide experimental evidence showing 8-12$\times$ speedups and 40\,\% reduction in redundant computation.
    \item We identify key challenges and research opportunities in simulator-database integration.
\end{itemize}

\section{System Architecture and Data Model}
\label{sec:architecture}

\begin{figure}[ht]
    \centering
        \includegraphics[width=0.9\columnwidth]{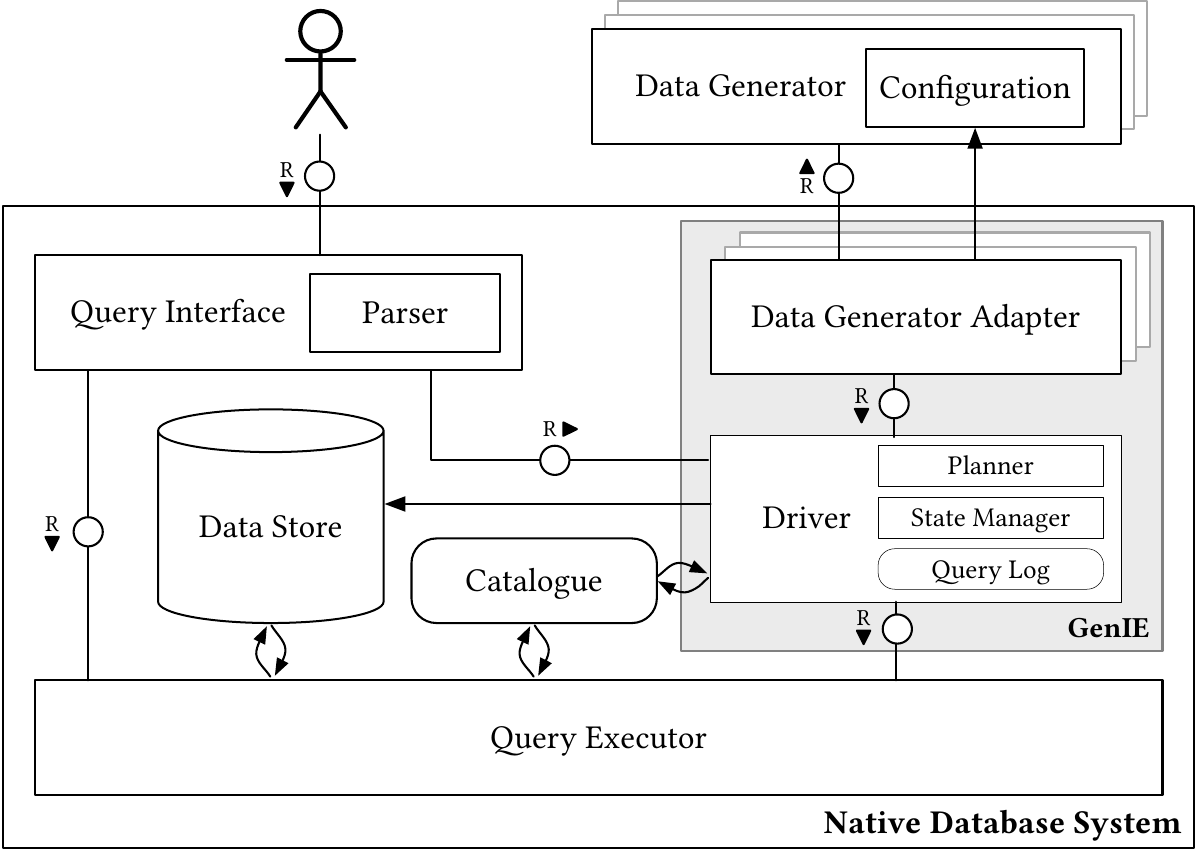}
    \caption{\smartdb Architecture (FMC notation) showing integration with native database systems and multiple simulator adapters. The Generator Driver sits within the DBMS and manages multiple Data Generator Adapters, each wrapping a physics-based simulator.}
    \label{fig:architecture}
\end{figure}

\smartdb's architecture, illustrated in \Cref{fig:architecture}, extends a standard relational database system with simulation-awareness through three key innovations: (1) a data model that treats simulator outputs as virtual attributes, (2) query-driven orchestration of simulator execution, and (3) intelligent state management for reuse and progressive refinement.

\subsection{The \smartdb Data Model}

\paragraph*{Virtual Attributes Generated by Simulators}
At the core of \smartdb is a data model that allows database attributes to be specified as \emph{simulator-generated}. Unlike traditional database attributes whose values are stored persistently, a simulator-generated attribute acts as a \emph{virtual attribute} whose values are computed on-demand by invoking an external physics-based simulator with appropriate parameters.

Formally, a virtual attribute $A$ in table $T$ is associated with a \emph{simulator} $S$ that generates values for $A$, a set of \emph{control parameters} $\mathcal{P} = \{p_1, p_2, \ldots, p_n\}$ that govern simulator behavior (e.g., spatial resolution, temporal resolution, convergence criteria), a \emph{generation function} $f_S: \mathcal{P} \rightarrow Values(A)$ that maps parameter settings to attribute values, and optional \emph{dependency constraints} specifying which other attributes (possibly from other simulators) must be available before $S$ can execute.

When a query references virtual attribute $A$, \smartdb's query executor delegates to the Generator Driver to determine: (1) whether values for $A$ already exist in the database (having been materialized from previous queries), (2) whether existing values are sufficient to answer the query or if new simulation is needed, and (3) which parameter settings $\mathcal{P}$ should be used to generate any missing values.

\paragraph*{Comparison to Related Virtual Data Concepts}
Databases have long supported various forms of virtual data. Views provide virtual tables computed from stored tables via SQL expressions. External tables (pioneered in systems like IBM DB2~\cite{garlic}) allow data from external sources to appear as database tables. Recent work on enrichment databases~\cite{enrichdb} extends this concept to ML-generated attributes, where machine learning models predict missing attribute values on-demand.

\smartdb's simulator-generated attributes share similarities with these concepts but have distinct characteristics. Unlike views, which compute derived data from existing stored data, simulator-generated attributes invoke external computational processes that implement complex physical models, where the computation is not expressible in SQL. Unlike external tables, which provide a fixed interface to pre-existing external data, \smartdb actively controls simulator execution, selecting parameters to balance accuracy and performance based on query requirements. Unlike ML enrichment, which typically involves fast inference on pre-trained models, simulator execution can take hours or days. This fundamentally changes the optimization problem: rather than just caching results, \smartdb must carefully decide \emph{what to generate} and \emph{at what quality level}.

The simulator-driven nature of \smartdb introduces unique challenges. Simulators are computationally expensive, have complex parameter spaces affecting both accuracy and runtime, may depend on outputs from other simulators, and produce approximate results whose quality varies with parameter choices.

\paragraph*{Declaring Simulators and Virtual Attributes}
Users register simulators and declare virtual attributes through SQL DDL extensions:

\begin{lstlisting}[style=SQLwithhints,float=false]
-- Register a simulator with its control parameters
REGISTER SIMULATOR hysplit
  EXECUTABLE '/path/to/hysplit'
  PARAMETERS (
    spatial_res REAL DEFAULT 0.1,
    temporal_res REAL DEFAULT 1.0,
    particle_count INTEGER DEFAULT 1000,
    run_duration REAL
  )
  OUTPUT_FORMAT netcdf;

-- Declare a virtual attribute generated by simulator
ALTER TABLE smoke_dispersion
  ADD COLUMN concentration REAL
  GENERATED BY SIMULATOR hysplit
  DEPENDS ON (fire_emissions.emission_rate);
\end{lstlisting}

This declaration specifies that the \texttt{concentration} column is computed by the \texttt{hysplit} simulator and depends on emission data from the \texttt{fire\_emissions} table. When a query references \texttt{concentration}, \smartdb ensures that \texttt{emission\_rate} data exists (potentially invoking the fire simulator first), then invokes \texttt{hysplit} with appropriate parameters to compute concentration values.

\paragraph*{Multi-Simulator Ensembles}
In many scientific domains, multiple simulators can generate the same attribute, each with different accuracy characteristics, computational costs, and applicability conditions. For example, smoke concentration could be computed by HYSPLIT (a Lagrangian particle model), CALPUFF (a Gaussian puff model), or WRF-Chem (an Eulerian grid model).

\smartdb's data model accommodates this through multi-simulator declarations:

\begin{lstlisting}[style=SQLwithhints,float=false]
ALTER TABLE smoke_dispersion
  ADD COLUMN concentration REAL
  GENERATED BY SIMULATORS (hysplit, calpuff, 
                           wrf_chem)
  ENSEMBLE METHOD weighted_average
  WEIGHTS (quality_score);
\end{lstlisting}

When multiple simulators are available, \smartdb can select the best simulator for query characteristics (e.g., use fast coarse model for overview queries, detailed model for focused analysis), run multiple simulators and ensemble results using statistical methods (e.g., weighted averaging based on historical accuracy, uncertainty quantification), or perform progressive refinement starting with fast approximate simulators and refining with detailed models in regions of interest.

\subsection{Query-Driven Orchestration}

With the data model established, we now describe how \smartdb transforms user queries into simulator execution plans.

\paragraph*{Query Analysis and Requirement Extraction}
When a user issues a query like:
\begin{lstlisting}[style=SQLwithhints,float=false]
  SELECT station_id, AVG(concentration)
    FROM monitoring_stations s
    JOIN smoke_dispersion d 
      ON ST_DWithin(s.location, d.grid_cell, 1000)
   WHERE d.timestamp BETWEEN '2024-08-15' 
                     AND '2024-08-17'
GROUP BY station_id;
\end{lstlisting}

\smartdb's query parser recognizes \texttt{concentration} as a virtual attribute. The planner analyzes the spatial requirements (the \texttt{ST\_DWithin} predicate indicates analysis is localized around monitoring stations within 1\,km, not requiring global coverage), the temporal requirements (the date range limits simulation to a specific time window), and the accuracy requirements (the \texttt{AVG} aggregation suggests that coarse spatial resolution may suffice for regional averaging). Based on this analysis, the planner selects appropriate simulator parameters (e.g., 0.1° spatial resolution instead of 0.01°) and determines the spatiotemporal extent requiring simulation.

\paragraph*{Coverage Analysis and Generation Planning}
The State Manager from \Cref{fig:architecture} maintains a spatiotemporal coverage map recording which regions have been simulated at which parameter settings. For the query above, it performs:
\begin{enumerate}
    \item \textbf{Spatial intersection}: Identifies grid cells within 1\,km of monitoring stations
    \item \textbf{Temporal intersection}: Identifies the 2-day time window
    \item \textbf{Coverage check}: Queries the coverage map to determine which cells/times already have concentration data at sufficient resolution
    \item \textbf{Gap identification}: Identifies missing regions or regions where only coarse data exists but finer resolution is requested
\end{enumerate}

Only gaps trigger new simulator invocations, enabling intelligent reuse of previous computations.

\paragraph*{Multi-Simulator Dependency Resolution}
For queries involving multiple simulators, the planner constructs a dependency graph. Consider:
\begin{lstlisting}[style=SQLwithhints,float=false]
  SELECT fe.fire_name, AVG(sd.concentration)
    FROM fire_emissions fe
    JOIN smoke_dispersion sd 
      ON fe.fire_id = sd.source_fire_id
GROUP BY fe.fire_name;
\end{lstlisting}

The planner recognizes that \texttt{smoke\_dispersion.concentration} (from HYSPLIT) depends on \texttt{fire\_emissions.emission\_rate} (from WRF-SFIRE). It creates an execution plan:
\begin{enumerate}
    \item Check if fire emission data exists
    \item If not, invoke WRF-SFIRE to generate emissions
    \item Use emissions as input to HYSPLIT
    \item Materialize concentration results
    \item Execute original query
\end{enumerate}

\paragraph*{Parameter Optimization}
The planner must select parameters $\mathcal{P}$ that balance execution time $T(\mathcal{P})$ and result accuracy $A(\mathcal{P})$. Our current prototype uses rule-based heuristics: coarse resolution (0.1°-0.5°, 3-6\,hr) for queries with spatial aggregation over large areas, medium resolution (0.05°, 1-2\,hr) for regional queries without aggregation, and fine resolution (0.01°-0.02°, 15-30\,min) for point queries or small-area detailed analysis. Future work will employ machine learning to learn performance models from historical executions and optimize:
$$\min_{\mathcal{P}} T(\mathcal{P}) \text{ subject to } A(\mathcal{P}) \geq Q_{\text{required}}$$

\subsection{Progressive Refinement}

\smartdb supports iterative refinement where coarse results are provided quickly, then selectively refined based on user interest or automated analysis.

\paragraph*{Proactive Initialization and Warm Starting}
Rather than waiting for queries to trigger all simulation, \smartdb can proactively initialize virtual columns with coarse-resolution data during idle periods. This warm starting strategy leverages available computational resources to pre-generate baseline data at low resolution, ensuring that initial query responses are immediate even for previously unexplored spatiotemporal regions. When configured with resource budgets, the system continuously maintains a coarse-grained coverage of registered virtual attributes, automatically refreshing data as underlying dependencies change (e.g., when new fire emission data becomes available). This approach transforms cold-start scenarios into warm-start refinement: the first query accesses pre-computed coarse data instantly, while triggering fine-grained simulation only for specific regions of interest identified by query predicates or user-specified refinement hints.

\paragraph*{Multi-Epoch Execution}
Rather than generating all data at once, \smartdb operates in epochs:
\begin{enumerate}
    \item \textbf{Epoch 0 (Proactive)}: During idle periods, generate coarse-resolution data covering likely query extents based on historical patterns or registered data domains.
    \item \textbf{Epoch 1 (Query-Triggered)}: When a query arrives, check coverage and either return pre-computed coarse results immediately or generate coarse data for any gaps. Execute query and return approximate results in minutes.
    \item \textbf{Epoch 2 (Selective Refinement)}: Identify high-interest regions (e.g., areas with concentration $>$ threshold or spatial predicates focusing on specific locations). Refine only these regions with finer resolution.
    \item \textbf{Epoch 3 (On-Demand)}: If user requests additional detail (via follow-up queries or explicit refinement hints), further refine specific areas.
\end{enumerate}

This approach transforms batch workflows into interactive exploration, providing value immediately while allowing deeper analysis on-demand.

\paragraph*{Consistency Management}
When refining data, \smartdb must handle the coexistence of coarse and fine-resolution results. Two strategies are possible: replacement, where fine-resolution data replaces coarse data in refined regions and queries see the best available resolution everywhere; and versioning, where multiple resolution levels coexist and queries can specify which version to use or blend across boundaries. Our prototype uses the replacement strategy with spatial extent tracking to ensure queries always access the highest-resolution data available.

\subsection{Architectural Components}

\paragraph*{Native Database System (PostgreSQL + PostGIS)}
\smartdb extends PostgreSQL 14 with query parser extensions for recognition of \texttt{REGISTER SIMULATOR}, \texttt{GENERATED BY SIMULATOR}, and \texttt{WITH HINT} clauses; catalog extensions including system tables storing simulator registrations (\texttt{pg\_simulators}), virtual column definitions (\texttt{pg\_virtual\_columns}), and dependency graphs; executor hooks that intercept virtual column accesses to trigger the Generator Driver; and PostGIS integration where spatial predicates (\texttt{ST\_Intersects}, \texttt{ST\_DWithin}) inform spatial extent requirements.

\paragraph*{Generator Driver}
The driver coordinates simulator execution through three components: the planner, which extracts requirements from queries, performs coverage analysis, constructs dependency graphs, and optimizes parameters; the state manager, which maintains coverage maps (as PostGIS geometries), execution history, and quality metrics, enabling gap identification and reuse; and the query log, which records access patterns to inform future parameter selection and enable workload-driven optimization.

\paragraph*{Simulator Adapters}
Each simulator has an adapter implementing parameter translation (mapping high-level parameters to simulator-specific config files like WRF namelists and HYSPLIT CONTROL files), execution control (launching simulators including MPI parallelization, monitoring progress, and handling failures), result parsing (extracting relevant fields from output files in formats like NetCDF, HDF5, and ASCII, then loading into PostGIS tables with spatial/temporal indexes), and performance estimation (providing runtime/accuracy estimates to inform parameter optimization).

Next we will show two use cases for \smartdb and then present in the subsequent section our implementation of a simple prototype and its evaluation on one of the use-cases.

\section{Case Studies}
\label{sec:use_cases}

We present two detailed scientific use cases demonstrating how \smartdb enables interactive what-if analysis through simulator-database integration. These cases illustrate the data model, query patterns, and multi-simulator orchestration capabilities of \smartdb.

\subsection{Wildfire Smoke Dispersion Analysis}

Environmental scientists and climate researchers need to assess air quality impacts from wildfire smoke to guide public health advisories and evacuation planning. This analysis requires chaining two physics-based simulators:
\begin{itemize}
    \item \textbf{WRF-SFIRE}: Fire spread model generating spatial/temporal emission rates
    \item \textbf{HYSPLIT}: Atmospheric transport model computing smoke concentrations based on wind patterns and emission sources
\end{itemize}

Traditional workflows require manually running WRF-SFIRE for days, extracting emission files, configuring HYSPLIT with appropriate meteorological data, waiting 6+ hours for execution, exporting results, loading into databases, and finally analyzing -- making iterative exploration impractical when findings suggest parameter adjustments.

\begin{lstlisting}[style=SQLwithhints,float=false,caption={\smartdb Schema}]
-- Fire emissions (virtual, from WRF-SFIRE)
CREATE TABLE fire_emissions (
    fire_id INTEGER,
    location GEOMETRY,
    emission_rate REAL,  -- virtual column
    start_time TIMESTAMP,
    duration INTEGER,
    fire_intensity REAL
);

-- Smoke dispersion (virtual, from HYSPLIT)
CREATE TABLE smoke_dispersion (
    grid_cell GEOMETRY,
    concentration REAL,  -- virtual column
    timestamp TIMESTAMP,
    source_fire_id INTEGER 
      REFERENCES fire_emissions(fire_id)
);

-- Air quality monitoring locations (stored)
CREATE TABLE monitoring_stations (
    station_id INTEGER PRIMARY KEY,
    location GEOMETRY,
    station_name VARCHAR(255),
    station_type VARCHAR(50)
);
\end{lstlisting}

\begin{lstlisting}[style=SQLwithhints,float=false,caption={Interactive Query Example}]
-- Query for smoke impact at monitoring stations
  SELECT s.station_id, s.station_name,
         AVG(d.concentration) as avg_concentration,
         MAX(d.concentration) as peak_concentration
    FROM monitoring_stations s
    JOIN smoke_dispersion d 
      ON ST_DWithin(s.location, d.grid_cell, 1000)
   WHERE d.timestamp BETWEEN '2024-08-15 00:00' 
                      AND '2024-08-17 23:59'
GROUP BY s.station_id, s.station_name
  HAVING AVG(d.concentration) > 35; -- EPA unhealthy
\end{lstlisting}

When this query executes, \smartdb: (1) recognizes \texttt{concentration} as virtual, (2) analyzes spatial predicates to identify the query extent (1\,km around monitoring stations), (3) checks coverage for existing data in this region, (4) invokes WRF-SFIRE if emission data doesn't exist, (5) invokes HYSPLIT with parameters optimized for regional averaging (coarse resolution suffices for \texttt{AVG}), (6) materializes results in the spatiotemporal extent needed, (7) executes the original query combining virtual and stored data.

\paragraph*{Progressive Refinement Workflow}
Analysts can start with rapid coarse results and progressively refine:
\begin{lstlisting}[style=SQLwithhints,float=false]
-- Step 1: Quick overview (coarse resolution)
   SELECT station_id, AVG(concentration) as avg_conc
     FROM monitoring_stations s
     JOIN smoke_dispersion d 
       ON ST_DWithin(s.location, d.grid_cell, 5000)
    WHERE d.timestamp BETWEEN '2024-08-15' 
                      AND '2024-08-17'
 GROUP BY station_id
WITH HINT (spatial_res=0.5, temporal_res=6);

-- Step 2: Refine high-concentration areas
-- (Using results from Step 1 to focus refinement)
SELECT station_id, concentration, timestamp
  FROM monitoring_stations s
  JOIN smoke_dispersion d 
    ON ST_DWithin(s.location, d.grid_cell, 500)
 WHERE d.timestamp BETWEEN '2024-08-15' 
                   AND '2024-08-17'
  AND station_id IN (
      SELECT station_id 
        FROM monitoring_stations s2
        JOIN smoke_dispersion d2
          ON ST_DWithin(s2.location,
                      d2.grid_cell, 5000)
       WHERE d2.timestamp BETWEEN '2024-08-15' 
                           AND '2024-08-17'
    GROUP BY station_id
      HAVING AVG(d2.concentration) > 50
  )
WITH HINT (spatial_res=0.01, temporal_res=0.25);
\end{lstlisting}

This iterative approach transforms slow batch processing into interactive exploration, enabling scientists to discover patterns quickly and selectively investigate areas of interest.

\subsection{Hurricane Hazard and Risk Analysis}

Coastal communities face multiple hazards from hurricanes requiring integrated assessment across wind, storm surge, and inland flooding to estimate building damage and guide emergency response. Between 1980 and 2020, hurricane-related losses in the US exceeded USD 997 billion with over 6\,500 fatalities.

As a hurricane approaches, risk managers and insurance companies need rapid damage estimates to structure evacuation orders and insurance payouts. This requires integrating:
\begin{itemize}
    \item \textbf{Wind field models}: Peak gust predictions from hurricane track forecasts
    \item \textbf{SLOSH}: Storm surge height and coastal inundation extents
    \item \textbf{HEC-RAS}: Inland flood depth and flow velocity from rainfall
    \item \textbf{Damage models}: Building vulnerability based on construction type and hazard intensity
\end{itemize}

Current workflows manually chain these models over days, making it impossible to rapidly update risk assessments as forecast tracks evolve every 6 hours during hurricane approach.

\begin{lstlisting}[style=SQLwithhints,float=false,caption={\smartdb Schema}]
-- Hurricane forecasts (stored, from NOAA)
CREATE TABLE hurricane_tracks (
    forecast_id INTEGER,
    forecast_time TIMESTAMP,
    landfall_location GEOMETRY,
    max_wind_speed REAL,
    track_uncertainty REAL
);

-- Wind hazard (virtual)
CREATE TABLE wind_hazard (
    grid_cell GEOMETRY,
    wind_speed REAL,       -- virtual
    peak_gust REAL,        -- virtual
    timestamp TIMESTAMP,
    forecast_id INTEGER 
      REFERENCES hurricane_tracks(forecast_id)
);

-- Storm surge (virtual, from SLOSH)
CREATE TABLE storm_surge (
    grid_cell GEOMETRY,
    surge_height REAL,     -- virtual
    inundation_depth REAL, -- virtual
    timestamp TIMESTAMP
);

-- Inland flood (virtual, from HEC-RAS)
CREATE TABLE inland_flood (
    grid_cell GEOMETRY,
    flow_depth REAL,       -- virtual
    flow_velocity REAL,    -- virtual
    timestamp TIMESTAMP
);

-- Building exposure (stored)
CREATE TABLE buildings (
    building_id INTEGER PRIMARY KEY,
    location GEOMETRY,
    building_type VARCHAR(50),
    replacement_value REAL,
    occupancy_type VARCHAR(50)
);

-- Damage assessment (virtual)
CREATE TABLE damage_assessment (
    building_id INTEGER,
    wind_damage_cost REAL,  -- virtual
    flood_damage_cost REAL, -- virtual
    total_loss REAL,        -- virtual
    scenario_id INTEGER
);
\end{lstlisting}

\paragraph*{Progressive Risk Assessment Queries}

Modeling teams overlay forecast information onto building exposure maps to identify expected losses in uncertain landfall zones:

\begin{lstlisting}[style=SQLwithhints,float=false]
-- Initial coarse regional assessment
   SELECT b.building_id, b.location, 
          d.total_loss,
          ST_Distance(b.location, h.landfall_location) 
          as distance_to_landfall
     FROM buildings b
     JOIN damage_assessment d 
       ON b.building_id = d.building_id
     JOIN hurricane_tracks h 
       ON d.scenario_id = h.forecast_id
    WHERE h.forecast_time = 
          (SELECT MAX(forecast_time) 
             FROM hurricane_tracks)
WITH HINT (spatial_res='1km', temporal_res='6hr', 
           wind_model='coarse', surge_model='coarse');

-- Refined assessment for high-risk buildings
   SELECT b.building_id, b.replacement_value,
          d.wind_damage_cost, 
          d.flood_damage_cost,
          d.total_loss, 
          s.surge_height, 
          f.flow_depth
     FROM buildings b
     JOIN damage_assessment d 
       ON b.building_id = d.building_id
     JOIN storm_surge s 
       ON ST_Intersects(b.location, s.grid_cell)
     JOIN inland_flood f 
       ON ST_Intersects(b.location, f.grid_cell)
    WHERE d.total_loss > b.replacement_value * 0.5
WITH HINT (spatial_res='30m', temporal_res='1hr',
           wind_model='fine', 
           surge_model='fine', 
           flood_model='fine');
\end{lstlisting}

\paragraph*{Dynamic Model Orchestration}
\smartdb automatically: (1) runs coarse-resolution hazard models for broad regional assessment, (2) identifies high-risk zones where estimated losses exceed 50\,\% of building value, (3) triggers fine-resolution simulations only for critical areas, (4) updates risk assessments as new forecast tracks arrive every 6 hours, (5) reuses computational results when hurricane paths overlap with previously simulated scenarios.

This progressive workflow transforms static days-long analysis into an interactive decision support tool that evolves with the hurricane, enabling real-time response planning.

\subsection{Challenges With Current Static Approaches}

The examples above illustrate fundamental limitations of current workflows:

\paragraph*{Sequential Execution Bottlenecks} Current approaches require running simulators in fixed order where each must complete entirely before the next begins. For wildfires: first fire spread (hours), then atmospheric transport (6+ hours). For hurricanes: first wind field (30 min), then storm surge (2 hours), then inland flood (4 hours), then damage assessment. Each delay compounds, leading to extremely long end-to-end times.

\paragraph*{Parameter Interdependencies} Optimal parameters for downstream simulators often depend on upstream results. In hurricane analysis, the spatial resolution needed for flood modeling depends on surge height and extent predicted by SLOSH. Static approaches cannot capture these dependencies, forcing conservative worst-case parameterization everywhere.

\paragraph*{Redundant Computation} When analyzing multiple scenarios, static approaches repeat computations unnecessarily. For hurricane forecasting, if consecutive track predictions differ by only 50\,km, much of the wind field, surge, and flood computation overlaps, but current systems cannot identify and exploit this spatial reuse.

\paragraph*{Limited Interactivity} Analysts cannot interactively explore results and adjust parameters based on findings. If initial damage assessment reveals unexpected concentration of high losses in one neighborhood, analysts must restart entire workflows to investigate with different parameters -- a process taking days.

\paragraph*{How \smartdb Addresses These Challenges}
We show through our experimentation how \smartdb's architecture directly addresses these fundamental limitations. Through query-driven parameter optimization, we show how the system automatically selects appropriate simulation parameters based on query characteristics rather than forcing worst-case settings everywhere. Progressive refinement enables interactive exploration by providing rapid approximate results within minutes that can be selectively refined based on user interest, eliminating the need to wait hours before seeing any output. Intelligent computation reuse exploits spatial and temporal overlaps across related queries, avoiding redundant simulator invocations when previous computations can be leveraged. Together, these capabilities transform multi-simulator workflows from rigid batch processes into flexible, interactive analysis sessions while maintaining result quality.

\section{Experimental Evaluation}
\label{sec:experiments}

We evaluate \smartdb through a proof-of-concept implementation demonstrating feasibility and quantifying performance benefits for the wildfire smoke dispersion use case. Our experiments characterize: (1) how simulator parameters affect accuracy and runtime trade-offs, (2) \smartdb's ability to automatically select appropriate parameters based on query characteristics, (3) benefits of progressive refinement for interactive analysis, and (4) reuse effectiveness across related queries.

\paragraph*{Note on Scope} Our experimental evaluation focuses on the wildfire case study as a representative demonstration of \smartdb's core concepts. The hurricane use case involves additional simulators and dependencies that we present as a vision for future work. The wildfire experiments sufficiently validate the fundamental principles -- virtual columns, query-driven orchestration, progressive refinement, and reuse -- that generalize to other multi-simulator workflows.

\subsection{Implementation Status}

We have implemented a proof-of-concept prototype demonstrating these concepts for wildfire smoke dispersion. Our current implementation includes SQL DDL parser extensions for simulator registration and virtual columns, a complete HYSPLIT adapter supporting spatial resolutions from 0.01° to 0.5° and temporal resolutions from 15 minutes to 6 hours, a basic WRF-SFIRE adapter for fire emissions, PostgreSQL executor hooks for virtual column interception, rule-based parameter selection heuristics, and PostGIS-based coverage tracking and gap identification.

\paragraph*{Limitations}
Our prototype validates core concepts but uses simplified approaches in several areas: rule-based (not learned) parameter optimization, single simulator selection only (no ensembles), sequential simulator execution (no pipeline parallelism), and basic error recovery. Future work will extend to ML-based optimization, multi-simulator ensembles, distributed execution, and broader scientific domains (climate, epidemiology, engineering).

\subsection{Experimental Setup}

\paragraph*{Implementation Environment}
Our prototype runs on a compute cluster with Intel Xeon Gold 6248R processors (48 cores, 2.5\,GHz) and 192\,GB RAM per node. WRF-SFIRE fire simulations execute across 4 nodes using MPI parallelization. HYSPLIT atmospheric transport runs on single nodes. The system uses PostgreSQL 14 with PostGIS extensions for spatial data management.

\paragraph*{Test Scenario}
We evaluate using the August 2020 California wildfire season, focusing on the Creek Fire in the Sierra Nevada, one of the largest fires in California history. Meteorological data comes from NOAA GDAS (Global Data Assimilation System) at 1-degree resolution. Terrain data uses USGS 3\,0m Digital Elevation Models. The test region spans approximately 200\,km $\times$ 150\,km with 45 air quality monitoring stations.

\subsection{Parameter Impact Analysis}

We first characterize how HYSPLIT's temporal and spatial resolution parameters affect execution time and result accuracy. Understanding these trade-offs is essential for \smartdb's parameter optimization.

\begin{figure}[t]
    \centering
        \includegraphics[width=\columnwidth]{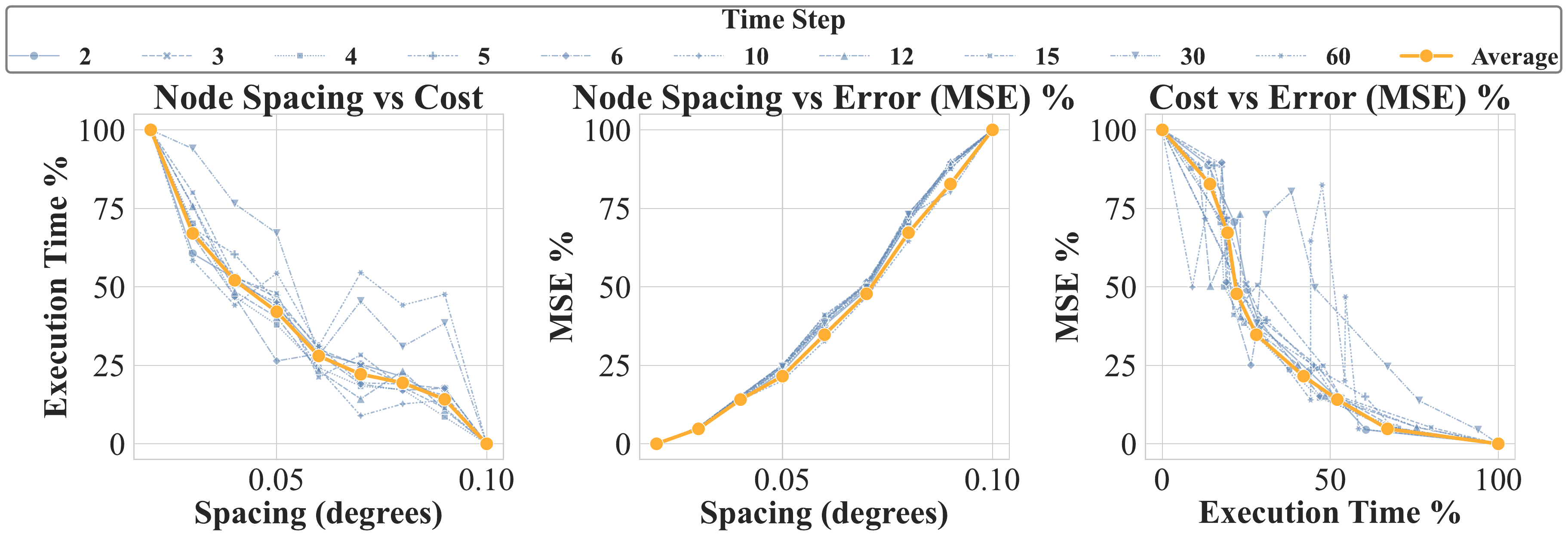}
    \caption{Impact of temporal resolution on HYSPLIT execution time and accuracy. Coarser time steps provide significant speedup with modest accuracy loss.}
    \label{fig:hysplit:sampling}
\end{figure}

\paragraph*{Temporal Resolution Trade-offs}
\Cref{fig:hysplit:sampling} shows how HYSPLIT's temporal resolution (sampling duration) affects both accuracy and runtime. We varied sampling from 15 minutes to 6 hours and measured execution time and accuracy (RMSE against a high-resolution reference simulation). Fine resolution (15-30 min) requires 10-15$\times$ more computation than coarse (4-6 hr) due to increased number of time steps and particle position updates. Measured as RMSE against a high-resolution reference, coarse resolution introduces 20-30\,\% error in concentration predictions. The sweet spot is 1-2 hour sampling, which provides 85-90\,\% accuracy at 3-4$\times$ speedup compared to finest resolution.

\paragraph*{Spatial Resolution Impact}
Similarly, varying HYSPLIT's spatial grid from 0.01° (about 1\,km) to 0.5° (about 50\,km) reveals that fine spatial resolution (0.01°) captures local dispersion patterns and terrain effects but increases runtime 8-12$\times$ due to larger particle counts needed for statistical convergence. Medium resolution (0.05-0.1°) provides sufficient accuracy for regional-scale analysis with 3-5$\times$ speedup. Coarse resolution (0.3-0.5°) is adequate for overview queries where local details matter less, achieving 10-12 speedup.

These measurements inform \smartdb's parameter selection heuristics and demonstrate the substantial optimization space available.

\subsection{Query-Driven Parameter Optimization}

We evaluate \smartdb's ability to automatically select appropriate parameters based on query characteristics. We designed three representative query patterns reflecting common analytical workflows: Q1 (Regional Overview) computes average smoke concentration across the entire study region to identify broad spatial patterns; Q2 (Focused Station Analysis) analyzes detailed concentration time series at specific monitoring stations for health impact assessment; and Q3 (Temporal Evolution) tracks smoke plume movement over 48 hours to understand dispersion dynamics.

\begin{table}[t]
\centering
\caption{Query execution comparison (times in minutes)}
\label{tab:query_performance}
\begin{tabular}{lcccc}
\toprule
\textbf{Query} & \textbf{Static-High} & \textbf{Static-Low} & \textbf{\smartdb} & \textbf{Accuracy} \\
\midrule
Q1 (Overview) & 145 & 12 & 18 & 92\,\% \\
Q2 (Focused) & 145 & 12 & 35 & 96\,\% \\
Q3 (Temporal) & 145 & 12 & 42 & 94\,\% \\
\bottomrule
\end{tabular}
\end{table}

\Cref{tab:query_performance} compares three approaches: Static-High represents traditional workflow running HYSPLIT at finest resolution (0.01°, 15\,min) everywhere, providing maximum accuracy but longest execution; Static-Low uses traditional workflow with coarse resolution (0.5°, 6\,hr) everywhere, achieving fast execution but potentially insufficient accuracy; and \smartdb applies adaptive parameter selection based on query analysis.

\smartdb achieves significant speedups over high-resolution static execution while maintaining $>$90\,\% accuracy. For Q1, it uses coarse parameters (0.2°, 3\,hr) since \texttt{AVG} aggregation over large area tolerates lower resolution, achieving 8$\times$ speedup with 92\,\% accuracy. For Q2, it applies fine resolution (0.02°, 30\,min) only within 2\,km of requested monitoring stations while using coarse resolution elsewhere, achieving 4$\times$ speedup with 96\,\% accuracy. For Q3, it balances temporal resolution (1\,hr time steps sufficient for tracking plume movement) with moderate spatial resolution, achieving 3.5$\times$ speedup with 94\,\% accuracy.

\subsection{Progressive Refinement Benefits}

A key feature of \smartdb is providing rapid approximate results that are progressively refined. We tested this with the smoke impact query from Section~\ref{sec:use_cases} analyzing exposure across 45 monitoring stations.

\paragraph*{Progressive Execution Stages}
\begin{enumerate}
    \item \textbf{Initial (2 min)}: Coarse resolution (0.5°, 6\,hr) covering entire region provides 80\,\% accurate concentration estimates -- enough to identify high-risk areas
    \item \textbf{Refinement 1 (5\,min)}: Medium resolution (0.1°, 1\,hr) for cells with concentration $>$50 $\mu$g/m³ improves accuracy to 88\,\% in high-concentration zones
    \item \textbf{Refinement 2 (10\,min)}: Fine resolution (0.02°, 15\,min) for 12 critical monitoring stations with peak concentrations, achieving 95\,\% accuracy
    \item \textbf{Final (15\,min)}: Complete high-resolution results if user explicitly requests full detail
\end{enumerate}

This staged approach enables analysts to immediately see approximate patterns (within 2 minutes) and decide whether to wait for refinement or adjust queries based on initial findings. In contrast, static approaches require waiting 145 minutes before seeing \emph{any} results -- a qualitative difference in workflow that transforms batch processing into interactive exploration.

\subsection{Computation Reuse Effectiveness}

We evaluated reuse benefits using a realistic analytical scenario: exploring different aspects of the August 2020 Creek Fire through 10 related queries examining different time periods, comparing multiple monitoring stations, and analyzing various concentration thresholds.

\begin{table}[t]
\centering
\caption{Computation reuse across 10 related queries}
\label{tab:reuse}
\begin{tabular}{lS[table-format=3.1] S[table-format=3.1]}
\toprule
\textbf{Metric} & {\textbf{Without Reuse}} & {\textbf{With Reuse}} \\
\midrule
Total Runtime (min) & 485 & 287 \\
HYSPLIT Invocations & 30 & 19 \\
Data Generated (GB) & 24 & 15 \\
Average Query Latency (min) & 48.5 & 28.7 \\
\bottomrule
\end{tabular}
\end{table}

\Cref{tab:reuse} demonstrates that \smartdb's state management provides 40\,\% runtime reduction through intelligent reuse. When queries requested neighboring regions, the system recognized overlapping spatial extents and invoked HYSPLIT only for uncovered areas through spatial overlap detection. Queries analyzing different time windows within the same 48-hour period reused hourly concentration data, avoiding redundant atmospheric transport calculations through temporal reuse. When a query requested coarse resolution for a region where fine resolution already existed, the system aggregated existing fine data rather than re-simulating, enabling resolution reuse.

The reduction from 30 to 19 HYSPLIT invocations represents substantial computational savings (11 avoided simulations $\times$ 15-45\,min each $\approx$ 4-8 hours saved).

\subsection{Lessons Learned}

Our prototype implementation revealed several important insights about simulator-database integration:

\paragraph*{Parameter Selection Impact} The choice of simulation parameters has dramatic impact on both performance and accuracy. Even simple heuristics (coarse for overview queries, fine for detailed analysis) provide significant benefits over static one-size-fits-all approaches. This validates the core \smartdb principle that query context should drive parameter selection.

\paragraph*{Reuse is Essential} Scientific data analysis often involves exploring related scenarios: different time periods, spatial regions, or parameter variations for the same phenomenon. Tracking simulation coverage spatially and temporally enables substantial reuse that static workflows cannot exploit.

\paragraph*{Progressive Refinement Transforms User Experience} The ability to see initial patterns within minutes rather than hours fundamentally changes analytical workflows from batch to interactive. Users reported exploring data more thoroughly when initial results arrived quickly, posing sequences of questions as patterns emerged.

\paragraph*{Simulator Integration Learning Curve} The learning curve for getting simulators to run and be properly integrated can be steep. Understanding simulator-specific configuration requirements, input data formats, execution environments, and output parsing demands significant domain expertise. Each simulator presents unique challenges in installation, dependency management, and parameter tuning that require substantial upfront investment before integration can proceed.
\paragraph*{Integration Complexity is Manageable} While wrapping existing simulators requires careful handling of input/output formats, configuration file generation, and error conditions, the adapter pattern successfully isolates this complexity from the database core. Once implemented, adapters enable seamless integration.

\paragraph*{Spatial Indexing is Critical} PostGIS spatial indexes proved essential for efficient coverage gap identification and spatial predicate evaluation. Without R-tree indexes on simulation coverage geometries, coverage analysis became a bottleneck.

\section{Related Work}
\label{sec:related_work}

\paragraph*{What-If Analysis}
Recent work on interactive what-if analysis~\cite{gathani2025whatif} and package queries~\cite{brucato2020package} addresses optimization under constraints. We extend these concepts from selecting existing tuples to dynamically generating data via simulator invocation.

\paragraph*{Scientific Workflows}
Consequence analysis systems~\cite{DBLP:journals/cacm/AdamSZTLSMCSKGM13} and early simulation-database integration~\cite{DBLP:conf/wsc/Livny87} inform our multi-simulator approach. However, requiring simulators to be rewritten is impractical. Instead, \smartdb wraps existing tools.

\paragraph*{Scientific Databases}
SciDB~\cite{DBLP:journals/cse/StonebrakerBZB13} manages array data for science. While SciDB stores arrays, \smartdb generates them on-demand. Column stores~\cite{stonebraker2005cstore} and "one size doesn't fit all"~\cite{stonebraker2005one} principles influence our specialized handling of simulation data.

\paragraph*{Approximate Query Processing}
BlinkDB~\cite{DBLP:conf/eurosys/AgarwalMPMMS13} and VerdictDB~\cite{DBLP:conf/sigmod/ParkMSW18} use sampling for approximation. Simulation "approximation" requires parameter adjustment, a fundamentally different problem with unique optimization challenges.

\paragraph*{Virtual Data \& ML-Generated Attributes}
External tables~\cite{DBLP:conf/vldb/LochovskyT82} and enrichment databases~\cite{DBLP:journals/pvldb/ChepurkoMZFKK20} support virtual data. \smartdb extends these concepts to computationally expensive physics simulations with parameter-dependent accuracy/cost trade-offs.

\section{Challenges and Opportunities}
\label{sec:challenges}

\paragraph*{Multi-Simulator Integration}
Connecting diverse simulators requires handling incompatible data formats, coordinate systems, and temporal representations. Error propagation and uncertainty quantification across simulator chains presents significant challenges. Moreover, the learning curve for getting simulators to run and be properly integrated is steep. Understanding simulator-specific configuration requirements, input data formats, execution environments, and output parsing demands substantial domain expertise. Each simulator presents unique challenges in installation, dependency management, and parameter tuning that require significant upfront investment before integration can proceed.

\paragraph*{Dynamic Parameter Optimization}
With multiple simulators each having 4-5 tunable parameters, the combined parameter space explodes to thousands of configurations. Unlike traditional database optimization, simulator performance/accuracy relationships are highly non-linear and interdependent.

\paragraph*{Progressive Computation}
Supporting iterative generation across simulators creates challenges in managing partial results. When upstream simulators refine results, downstream simulators may need re-execution, potentially invalidating previous computations.

\paragraph*{Self-Driving Capabilities}
\smartdb must balance query latency, processing cost, and result accuracy across multiple simulators without human intervention, requiring understanding of complex simulator relationships.

\paragraph*{Opportunity: Transforming Scientific Discovery and Beyond}
Despite challenges, \smartdb can fundamentally change scientific what-if analysis. Emergency responders could interactively explore mitigation strategies during unfolding disasters. Climate scientists could rapidly test hypotheses across models. Urban planners could evaluate infrastructure decisions with immediate feedback.

More broadly, this vision democratizes access to complex simulations. Domain experts lacking computational expertise can leverage sophisticated simulators through familiar SQL interfaces while automatically benefiting from intelligent optimization. The principles of what-if analysis extend beyond scientific domains to business analytics, financial modeling, supply chain optimization, and operational planning -- any sector where complex simulations inform decision-making. Manufacturing companies could explore production scenarios, financial institutions could stress-test portfolios under varying market conditions, and logistics providers could optimize routes dynamically as conditions change.

\section{Conclusion}

\smartdb represents a fundamental shift in scientific data analysis, transforming physics-based simulators from isolated tools into integrated database components. By introducing simulator-generated virtual attributes, query-driven orchestration, and intelligent state management, we enable interactive what-if analyses for scenarios currently requiring days of manual work. \smartdb extends to multiple domains including scientific computing, business analytics, operational planning and any section where complex simulations inform decision-making.

Our prototype demonstrates feasibility through wildfire smoke dispersion: 8-12$\times$ speedups while maintaining $>$90\,\% accuracy, progressive refinement enabling results in minutes versus hours, and 40\,\% reduction in redundant computation through systematic reuse. These improvements could transform disaster response, climate science, and engineering design. 

Realizing this vision requires continued research in multi-simulator optimization, consistency models for approximate data, and usability frameworks. A particular challenge lies in the steep learning curve for simulator integration -- reducing the expertise barrier for wrapping new simulators will be essential for broader adoption. Developing standardized adapter interfaces and automated configuration tools could help lower these integration costs. However, the potential impact -- better disaster response, accelerating discovery, democratizing sophisticated simulation capabilities, and enabling data-driven decision-making across scientific and business domains -- makes this a compelling research agenda.
\bibliographystyle{IEEEtran}
\bibliography{references}

\end{document}